# Phase Transformations During Continuous Cooling in Inconel 718 Alloys Manufactured by Laser Powder Bed Fusion and Suction Casting

**Yunhao Zhao[1], Liangyan Hao[1], Qiaofu Zhang[2], Wei Xiong[1]***

[1]Physical Metallurgy and Materials Design Laboratory,
Department of Mechanical Engineering and Materials Science,
University of Pittsburgh, Pittsburgh, PA 15261, USA
[2]QuesTek Innovations LLC,
1820 Ridge Avenue, Evanston, IL 60201, USA
* Corresponding Author, Email: weixiong@pitt.edu, w-xiong@outlook.com
Tel. +1 (412) 383-8092, Fax: +1 (412) 624-4846

## Abstract

Understanding alloy phase transformations during continuous cooling is important for post-processing design and optimization. In this work, continuous-cooling-transformation (CCT) diagrams of Inconel 718 alloys manufactured by laser powder bed fusion (LPBF) and suction casting are developed under different homogenization conditions. Unlike the available CCT diagrams in the reported studies, no $\gamma''$ and $\gamma'$ precipitates can be observed. NbC and $\delta$ are determined to be the precipitates after cooling from the $\gamma$ matrix. Importantly, homogenization time and manufacturing methods are found to affect the Nb homogeneity in the matrix near NbC particles and thus significantly influence the precipitation process of the $\delta$ phase, which has a high content in Nb. In the alloys with high Nb homogeneity, the nucleation process mainly contributes to the precipitation, whereas in the alloys with low Nb homogeneity, the precipitation is primarily associated with the growth process. Subgrains are found to form after cooling at 0.1 K/s and can cause the highest hardness in samples. This work provides a new viewpoint on the study of processing-structure-property relationships during cooling in Inconel 718 and is beneficial to the development of alloy post-processing strategies.

## 1. Introduction

Continuous-cooling-transformation (CCT) diagrams of alloys have been used very often to study solid-solid phase transformations. Since the CCT diagrams are conducive to understanding the microstructure-property relationships during processing and thus further provide design solutions for microstructure engineering. Due to the importance of the Inconel 718 in engineering applications with good high-temperature performance [1–3], the evaluation of CCT diagrams becomes critical, especially for additive manufacturing (AM) such as laser powder bed fusion (LPBF). In recent years, the application of Inconel 718 in AM has also proliferated owing to its good weldability as well as the increasing demands for the high-temperature components with complex shapes [4–6]. Therefore, it is essential to understand the phase stability and phase transformations under different heat treatment and cooling conditions.

Inconel 718 is an fcc precipitation-hardenable superalloy strengthened by the γ″ ($Ni_3Nb$, bct_$D0_{22}$) plates and the spherical γ′ ($Ni_3$(Al, Ti, Nb), fcc_$L1_2$) particles [7,8]. Besides, NbC carbides (fcc_B1) usually form along with γ at high temperatures. The δ ($Ni_3Nb$, $D0_a$) and Laves_C14 ($(Ni, Fe, Cr)_2$(Nb, Ti, Mo), hexagonal) are two detrimental intermetallic phases that are often observed in the alloy. The δ phase has a needle shape and usually precipitates along grain boundaries [8,9], and thus potentially can act as a pinning particle for grain size control but degrades strengthening property due to its incoherent phase boundary with the γ matrix [10]. The Laves_C14 phase forms during solidification as a result of Nb segregation at grain boundaries. In order to avoid crack initiation due to segregation, Inconel 718 alloys are often homogenized at elevated temperature as one of the most effective solutions [11–13].

It has been found that the phase transformation behaviors in Inconel 718 strongly depend on the applied heat treatments [13–16]. Therefore, understanding phase transformations during the continuous cooling process is critical for the microstructure engineering. This is particularly important for the AM process, in which the cyclic heating and cooling processes introduce complexity in the microstructure evolution, and thus usually require careful design of both in-situ processing and post-heat treatments [10,14,17–22]. CCT diagrams directly provide the information regarding phase transformations during the continuous cooling processes and are often applied as a tool together with isothermal transformation diagrams for microstructure engineering. Such a diagram can be useful in post-heat treatment but can also be applied to get insight during the cyclic heating and cooling processes when only solid phases are involved.

A few CCT diagrams of Inconel 718 have been previously reported [14,23–25], as summarized in Fig. 1. In order to facilitate discussion, we define three characteristics of the CCT curves, as illustrated in Fig. 1(a), to describe the phase transformation behaviors: (1) starting temperature of the phase transformation; (2) phase formation range, which is defined as the difference between the starting temperature and ending temperature at one certain cooling rate; and (3) critical cooling rate, below which the phase transformation will happen. Garcia et al. [14] used a dilatometer to investigate the effect of homogenizations at 1180°C for 24, 72, and 90 h on the CCT diagrams of cast Inconel 718 alloys. They found the CCT curves shifted to the slower reaction side, i.e., the right-hand-side of the CCT diagram, with increased homogenization durations (Figs. 1(b)-(f)). The authors [14] also reported that during cooling after 90-h homogenization, new and small MC carbides formed before the δ phase, which was different from the cases of 24 and 72-h homogenizations. This was explained as the Nb segregation along grain boundaries was reduced after long-time homogenization, which increased the Nb supersaturation within the grains and promoted the formation of MC carbides, but limited δ. However, the formation temperature of the δ phase was determined to be quite high (Fig. 1(c)), which is up to about 1130°C. This temperature is much higher than the reported solvus temperatures of δ from 998.3 to 1027°C obtained from experiments and CALPHAD (calculation of phase diagrams) calculations [26–29]. In addition, the γ″, γ′, and Laves_C14 phases were also found to form after continuous cooling, as can be seen in Figs. 1(d),(e)&(f). γ″ was determined to precipitate prior to γ′ during cooling. The critical cooling rates for γ″ were 1-10 K/s, depending on the homogenization time (Fig. 1(d)). Geng et al. [23] investigated the phase transformation behaviors of γ″ and δ during continuous cooling after homogenization at 1100°C for 1 hour in hot-extruded Inconel 718 alloys. The result deviated significantly from the one by Garcia et al. [14] since they determined the γ″ precipitated under very slow cooling rates of 0.1~20 K/min (0.0017-0.33 K/s), while δ formed at cooling rates lower than 5 K/min (0.083 K/s). Slama and Cizeron [24] reported that the δ, γ′, and γ″ phases can precipitate respectively after heat treatment at 990°C for 15 min, as reproduced in Fig. 1(d). The critical precipitation cooling rate of the δ phase was determined to be higher than 100 K/s, which was extremely high compared to the values from other work; while for the γ′ and γ″, critical cooling rates were 5 K/s, and 0.2 K/s, respectively. Niang et al. [25] provided a CCT curve for δ measured by differential thermal analysis (DTA) in forged Inconel 718 alloys, the critical cooling rate of δ was evaluated to be about 0.5 K/s (Fig. 1(c)). These results show that the homogenization conditions can affect the phase transformation behaviors during continuous cooling.

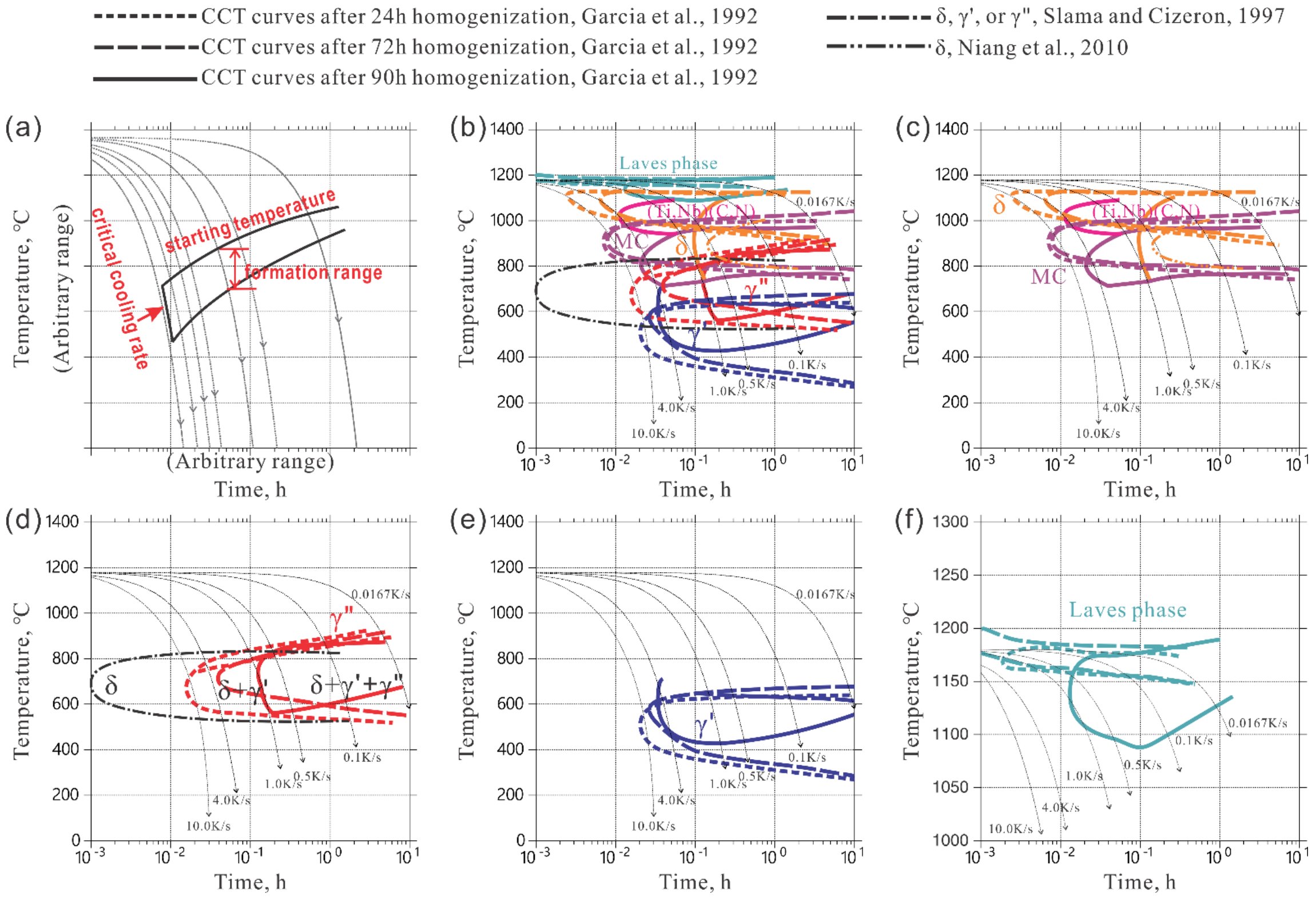

*Figure 1. (a) Defined characteristics of CCT diagrams in this work: starting temperature, phase formation range, and critical cooling rate; (b) integration of the reported CCT diagram by Garcia et al. [14], Slama and Cizeron [24], and Niang et al. [25]. Individual phases plotted in (b) are shown in the separate plots from (c) to (e) to show each section more clearly; (c) separate plot of the CCT diagram related to the δ phase, carbides, and carbonitrides according to Refs. [14,25]; (d) separate plot of the CCT diagrams related to γ″ phase (in red curves) according to Garcia et al. [14] and related to the δ, γ′, or γ″ phases (in black curves) based on Slama and Cizeron [24]; (e) separate plot of the CCT diagram related to the γ′ phase according to Garcia et al. [14]; (f) separate plot of the CCT diagram related to the Laves phase according to Garcia et al. [14]. The different line types from Garcia et al. [14] represent the CCT curves determined after various homogenization times.*

Although some CCT diagrams of cast/wrought Inconel 718 have been reported, the results from different work are inconsistent and dependent on the alloy fabrication status. Moreover, it is yet unclear regarding the impact of the manufacturing method on the CCT diagrams, since processing and heat treatment design is often based on the reported diagrams of the cast Inconel 718. However, as indicated by Zhao et al. [13], the as-received and homogenized microstructure of Inconel 718 manufactured by LPBF and suction-cast are significantly different. The as-built microstructure after LPBF shows a strong texture with columnar grain along build direction with a low Nb microsegregation. During homogenization at 1180°C, the columnar grains in the as-built alloy become equiaxial due to recrystallization and grain refinement occurs as a result of the Zener pinning effect. The Nb homogeneity level in the γ matrix decreases during homogenization. Contrarily, the grain morphology in the as-cast Inconel 718 is equiaxial and the Nb segregation level is high. Moreover, abnormal grain growth can be observed in the suction-cast alloys during homogenization at 1180°C and the Nb homogeneity level increases. Therefore, the microstructures of LPBF and suction-cast alloys after homogenizations are quite different, indicating the manufacturing methods can also impact the phase transformations of Inconel 718 during continuous cooling processes after homogenization and such effect is worth of more dedicated study.

This work aims at a comprehensive evaluation of the phase transformation behaviors during continuous cooling processes of Inconel 718 alloys under different homogenization conditions and manufacturing methods. Alloys made by both LPBF and suction casting are subject to microstructure analysis and quenching dilatometry. The suction-cast alloys are chosen as a reference for comparison because they have comparative phase transformation behaviors during homogenization processes from AM alloys [13]. Microhardness of alloys after cooling is studied to help understand the microstructure-property correlations under different cooling conditions.

## 2. Experiments and Computation

### 2.1. Experiments

The AM alloys were printed by an EOS M 290 machine using default build parameters, which are optimized for Inconel 718 by the EOS company. The build parameters can be found in [13]. The suction-cast alloys were made into small cylinders with a 40 mm-length and a 11 mm-diameter through an ABJ-338 arc-melter made by Materials Research Furnace Inc. under an Argon atmosphere to avoid oxidation. The nominal compositions of these two alloys are close, with Ni-18.26Fe-18.87Cr-4.97Nb-2.97Mo-0.94Ti-0.46Al-0.03C-0.06Mn-0.23Co-0.05Cu-0.06Si (in wt.%) for the AM alloys, and Ni-18.5Fe-18.3Cr-4.99Nb-3.04Mo-1.02Ti-0.55Al-0.051C-0.23Mn-0.39Co-0.07Cu-0.08Si (in wt.%) for the suction-cast alloys, respectively. Both alloys were sectioned by EDM wire cutting into cuboids with a dimension of 4×4×10 mm for dilatometry measurements. The length (10 mm) of the cuboids of AM alloys is along their build direction.

**Table 1. Sample notations and heat treatment conditions in the present work.**

| AM samples (LPBF) | | | Suction cast samples | | |
|---|---|---|---|---|---|
| **Sample notation*** | **Homogenization** | **Cooling rate, K/s** | **Sample notation** | **Homogenization** | **Cooling rate, K/s** |
| AM20m-01 | 1180°C for 20 min | 0.1 | AC20m-01 | 1180°C for 20 min | 0.1 |
| AM20m-1 | | 1 | AC20m-1 | | 1 |
| AM20m-2 | | 2 | AC20m-2 | | 2 |
| AM20m-5 | | 5 | AC20m-5 | | 5 |
| AM20m-7 | | 7 | AC20m-7 | | 7 |
| AM20m-10 | | 10 | AC20m-10 | | 10 |
| AM20m-15 | | 15 | AC20m-15 | | 15 |
| AM12h-01 | 1180°C for 12 h | 0.1 | AC12h-01 | 1180°C for 12 h | 0.1 |
| AM12h-1 | | 1 | AC12h-1 | | 1 |
| AM12h-2 | | 2 | AC12h-2 | | 2 |
| AM12h-5 | | 5 | AC12h-5 | | 5 |
| AM12h-7 | | 7 | AC12h-7 | | 7 |
| AM12h-10 | | 10 | AC12h-10 | | 10 |
| AM12h-15 | | 15 | AC12h-15 | | 15 |

* *AM20m is used to note a set of alloys with the same homogenization (1180°C for 20 min) but different cooling rates, and thus includes seven samples: AM20m-01, AM20m-1, AM20m-2, AM20m-5, AM20m-7, AM20m-10, and AM20m-15. The same way of notation applies to other alloys.*

The cuboid samples were subject to dilatometry with an S-type thermocouple. For each alloy, two groups of samples were further divided with respect to different homogenization conditions, i.e., 1180°C for 20 min and for 12 h, respectively. Samples of each group were then put into a quenching dilatometer (DIL805A, TA company) for the homogenization and subsequent cooling processes to room temperature. The temperature measurement error limit is ±1°C in this study. The cooling rates were 0.1, 1, 2, 5, 7, 10, and 15 K/s. The sample notations and heat treatment conditions in the present work are summarized in Table 1. It should be pointed out that, the samples to be homogenized for 12 h were firstly encapsulated into quartz tubes with backfilled Argon and were then homogenized into the furnace at 1180°C for 11 h. These 11-h homogenized samples were then quenched into ice-water and put into the dilatometer for the remaining 1-h homogenization at 1180°C. This can improve the efficiency of the heat treatment and avoid a long-time experiment in the dilatometer. The continuous cooling processes with different rates were conducted on these samples subsequently.

By checking the similarity of the obtained dilatation curves, typical as-cooled samples were selected from each group for microstructure characterization. These samples were polished using metallographic methods for SEM (scanning electron microscopy, Zeiss Sigma 500 VP, Carl Zeiss AG) and EDS (electron dispersive spectroscopy, Oxford Instruments plc) characterizations. The samples were also etched using a solution of 50 mL $C_3H_6O_3$+30 mL $HNO_3$+2 mL HF to reveal the existence of Nb-rich γ″ and γ′ phases [30]. Vickers hardness testing was performed using a Leco LM-800 tester under a 50 gf load for 10 s.

### 2.2. Model-prediction of phase stability

In order to understand the phase stability in Inconel 718, the equilibrium and nonequilibrium step diagrams of Inconel 718 are plotted by Thermo-Calc software using the TCNI8 thermodynamic database, as shown in Figs. 2(a)&(b), respectively. Since the γ″ phase is a metastable phase and can transform into the δ phase after long-time aging processes [31,32], the nonequilbrium step diagram of Inconel 718 was calculated by suspending the formation of the δ phase to predict the formation of γ″, whereas all other calculation parameters are the same with the equilibrium step diagram. According to the model-prediction, the NbC carbide is stable up to 1300°C and can co-exist with γ during homogenization at 1180°C. δ and γ′ are the stable phases from Fig. 2(a).The calculated step diagrams can guide the precipitation analysis from dilatometry measurements.

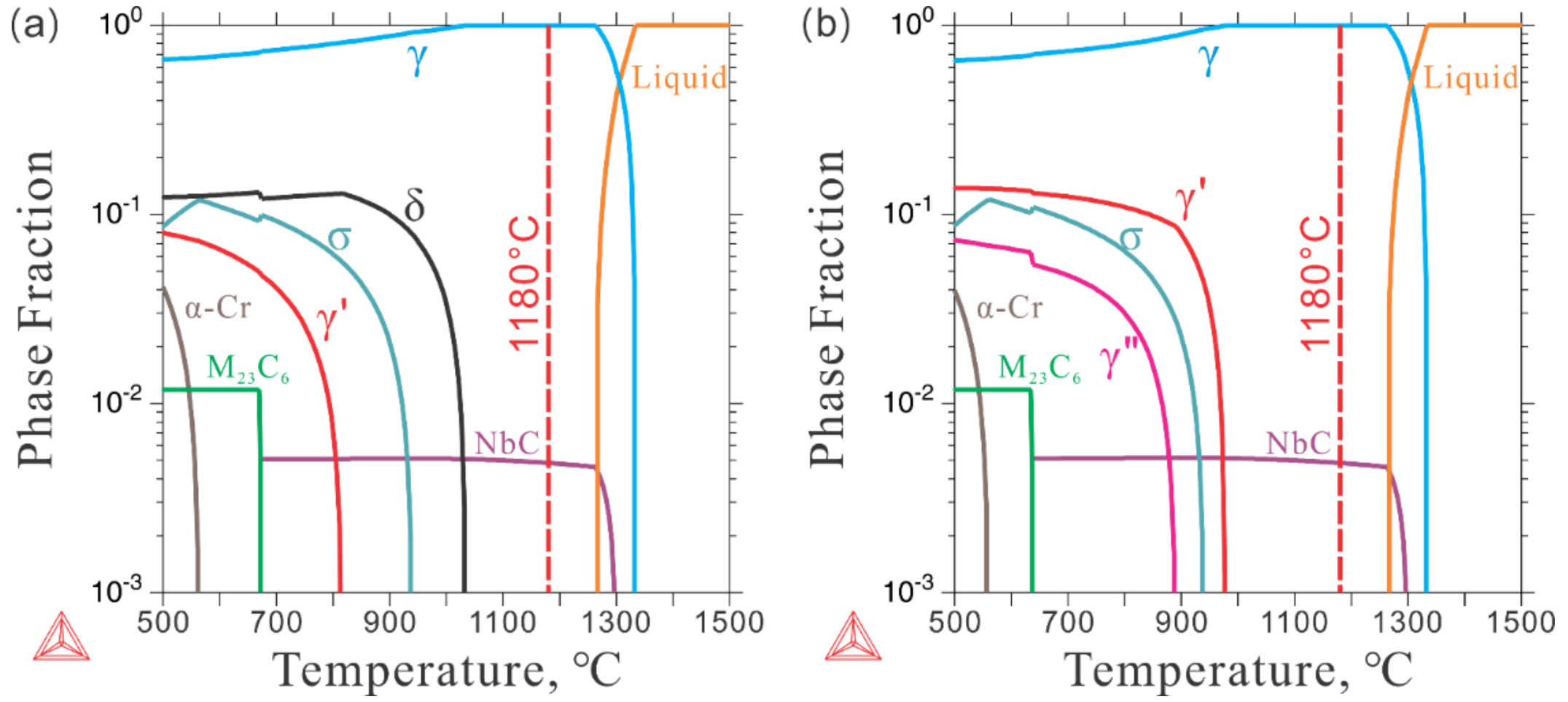

*Figure 2. Phase stability prediction according to the Thermo-Calc database TCNI8. (a) Equilibrium step diagram of Inconel 718; (b) nonequilibrium step diagram of Inconel 718.*

### 2.3. Simulation of precipitation kinetics

As will be discussed later in section 4, the δ phase is found to be the main precipitate during continuous cooling and its precipitation is impacted by Nb homogeneity in the matrix. In order to understand the underlying physics of the δ phase formation observed in experiments of this work, precipitation simulation is performed to qualitatively validate the effects of Nb homogeneity in the matrix near NbC carbides on the CCT diagrams. The simulation is conducted on the TC-PRISMA module implemented in the Thermo-Calc software using the TCNI8 and MOBNI4 databases. The simulation parameters are summarized in Table 2, where a lower Nb content of 4.99 wt.% (i.e., 3.14 at.% Nb) and a higher nucleation site number of 3E22 $m^{-3}$ are used to simulate the condition of high Nb homogeneity similar to sample AM20m; whereas a higher Nb content of 6.00 wt.% (i.e., 3.79 at.% Nb) and a lower nucleation site number of 3E21 $m^{-3}$ are used to simulate the condition of low Nb homogeneity similar to the case of alloy AM12h. It is noteworthy that the Nb contents for simulation (3.14 at.% and 3.79 at.%) were arbitrarily chosen to represent the experimentally determined Nb contents of 3.0-5.1 at.% as shown in Fig. 6. This will ensure the simulation results to be representative for more prevalent circumstances. A model system Ni-Fe-Cr-Nb-Mo instead of the actual Inconel 718 alloy is considered to improve the simulation efficiency. This will still gain enough insights into the multiphysical influence, particularly Nb homogeneity, on the CCT diagram of Inconel 718 under different homogenization conditions. The detailed discussion about the simulation results will be given in section 4.3.3.

**Table 2. Parameters of the δ phase precipitation simulation for validating the effects of Nb homogeneity.**

| Conditions | Composition, wt.% | Nucleation sites, /m$^3$ |
|---|---|---|
| High Nb homogeneity (e.g., AM20m) | Ni-18.26Fe-18.87Cr-2.97Mo -4.99Nb | 3E22 |
| Low Nb homogeneity (e.g., AM12h) | Ni-18.26Fe-18.87Cr-2.97Mo -6.00Nb | 3E21 |

## 3. Key equations in classical nucleation and growth theory

The classical nucleation theory is applied in this work to gain insights into the mechanisms of phase transformations during cooling processes. The determination of the nucleation rate $N_r$ in this work is based on the classical nucleation theory [33,34]. Assume the nucleation process is steady, $N_r$ is expressed as

$$N_r = Z\beta^* N_0 \exp\left(-\frac{\Delta G^*}{k_B T}\right) \qquad (1)$$

where $Z$ is the Zeldovich factor, which gives the probability of a nucleus to form a new phase, $\beta^*$ is the attachment rate of atoms to the critical nucleus, $N_0$ is the number of potential nucleation sites of the δ phase, $\Delta G^*$is the nucleation barrier, $k_B$ is the Boltzmann constant, $T$ is temperature. From the expressions of $Z$ and $\beta^*$ [35], at a certain temperature and given a certain phase in an alloy, the two factors can be considered as constants. Hence, according to Eq. (1), both increased $N_0$ and decreased $\Delta G^*$can enhance the nucleation rate $N_r$.

In addition to the nucleation rate, $N_r$, the growth rate can also affect the precipitation kinetics. The growth rate of the δ particle $J_r$ is expressed by [36]:

$$J_r = \frac{\Delta X_0^{Nb}}{2\left(X_\delta^{Nb} - X_e^{Nb}\right)}\left(\frac{D}{t}\right)^{\frac{1}{2}} \qquad (2)$$

of which $\Delta X_0^{Nb} = X_0^{Nb} - X_e^{Nb}$ is the supersaturation of Nb in the matrix, $X_0^{Nb}$ is the Nb concentration in the matrix, $X_e^{Nb}$ is the Nb equilibrium concentration in the matrix adjacent to the δ particles. $X_\delta^{Nb}$ is the Nb concentration in equilibrium δ phase, $D$, and $t$ are the interdiffusion diffusivity of Nb and diffusion time, respectively. Since the δ phase is an intermetallic with the formula of $Ni_3Nb$, and local equilibrium in the interface is assumed [36], the concentrations of Nb in the δ phase and the γ matrix can be considered as constants. Therefore, the diffusivity of Nb, $D$, is independent of composition and can be regarded as a constant under a certain temperature. According to Eq. (2), it is deduced that when the supersaturation $\Delta X_0^{Nb}$ increases, the growth rate $J_r$ is increased.

During phase transformation analysis in this work, we apply the above equations to comprehend how the Nb homogeneity in the matrix around NbC carbides influence the nucleation site $N_0$, nucleation barrier $\Delta G^*$, the supersaturation $\Delta X_0^{Nb}$ to determine their effects on precipitation kinetics.

## 4. Results and discussions

### 4.1. Microstructure characterization and dilatometry analysis

The CCT diagrams are determined through the combined microstructure and dilatometry analysis. The microstructure was characterized to investigate the phase transformations that occur during cooling, and the signals obtained from dilatation curves can be interpreted accordingly. The analysis of sample AM12h-5 is taken as an example, as shown in Fig. 3. The SEM micrograph (Fig. 3(a)) on the longitudinal plane parallel to the build direction of sample AM12h-5 shows block-shaped NbC carbides and needle-shaped δ precipitates form along the grain boundaries of γ matrix. Neither γ″ nor γ′ phase is observed in the etched sample AM12h-5, as their precipitation was suppressed by the fast cooling.

The starting and ending temperatures of phase transformations during continuous cooling are determined by the dilatometry analysis. As indicated by Fig. 3(b), the dilatation is plotted as a function of temperature. The slope changes in the dilatation curve represent the occurrence of phase transformations. Because of the small fraction (less than 1% in total) of the precipitates formed during cooling processes, the slope change of dilatation in this work is small, which may reduce the accuracy of the measurement of phase transformation temperature. Hence, the phase transformation temperatures are evaluated to correlate the inflection points which represents the slope change of the dilatation curve. The inflection points locate in where the first derivative of the dilatation curve reaches the maximal/minimal values, which can be determined by solving the zero points of the second derivative of the dilatation curve [37]. From Figs. 3(c)-(e), it can be seen from the 1st derivative curves of the dilatation curve of sample AM12h-5 that there are two inflection points for each curve segment indicating the NbC precipitation ending point, the δ precipitation starting point, and the δ precipitation ending point, respectively. It should be noted that, from Fig. 2(a), as the NbC is stable at 1180°C with γ matrix, the formation of NbC carbides should have started during homogenization processes at 1180°C. The starting temperature of NbC carbides during the cooling process is not available, and only the ending temperature can be determined. Therefore, it can be deduced that for the starting of precipitation during the cooling process, the inflection point at higher temperature should indicate the phase transformation point since the dilatation slope should keep constant before the phase transformation happens; on the contrary, for the ending point of precipitation, the transformation point should be indicated by the inflection point at the lower temperature, because the dilatation slope should be a constant after the ending of the precipitation.

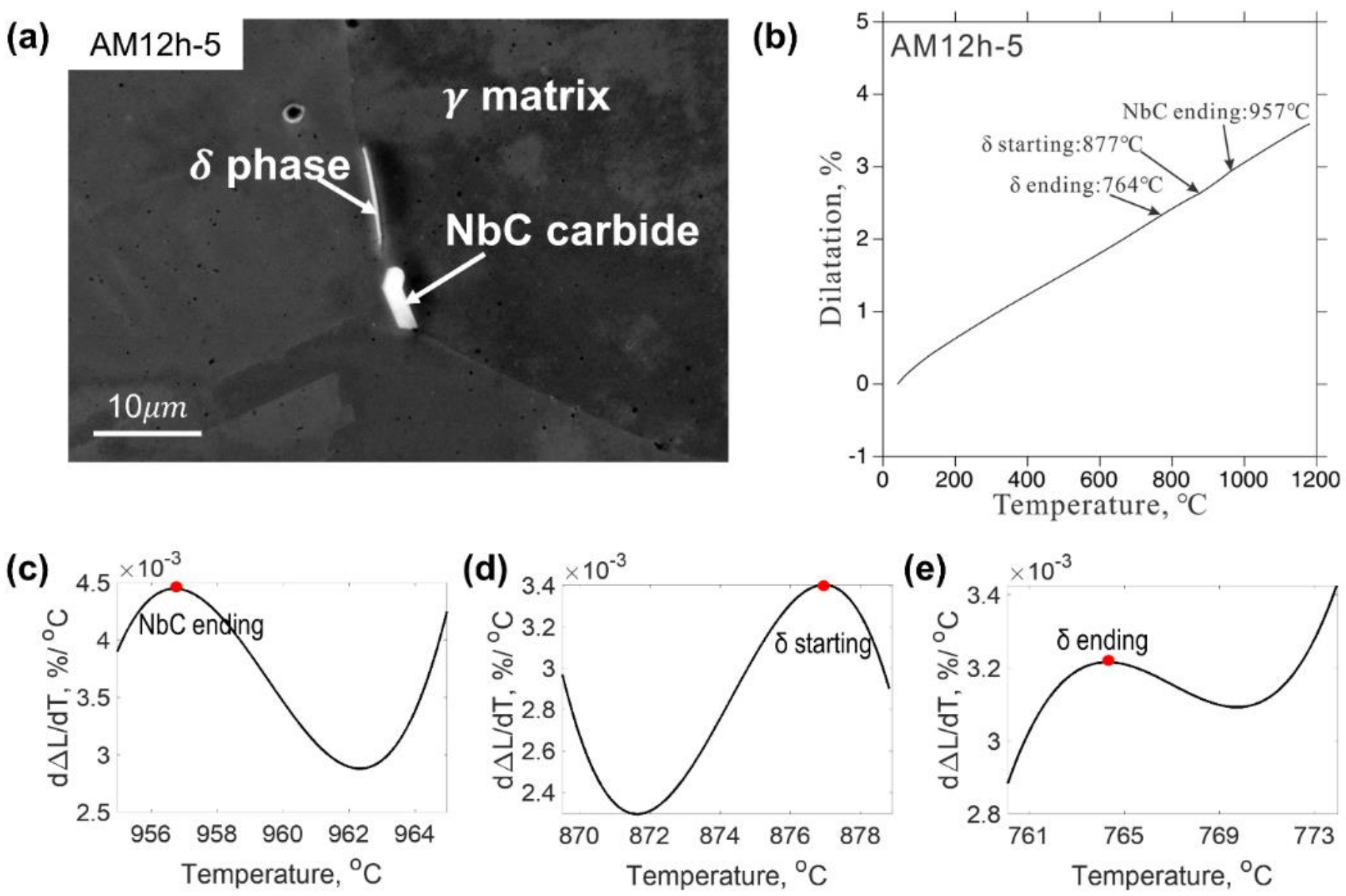

*Figure 3. Microstructure characterization and dilatometry analysis of sample AM12h-5. (a) SEM micrograph on the longitudinal plane of AM12h-5. (b) Dilatation curve of sample AM12h-5 and its first derivative curves for determining the inflection points of phase transformations for (c) NbC ending; (d) δ starting; and (e) δ ending.*

In sample AM12h-5 (Fig. 3(b)), the ending temperature of NbC precipitation is determined to be 957°C at which the second slope change occurs. Similarly, the first change of the slope at 877°C is determined as the starting temperature of δ precipitation. The ending temperature of δ precipitation is determined as 764°C. The phase transformation temperatures for the rest samples experiencing continuous cooling are determined using the same methods.

### 4.2. CCT diagrams of AM and suction-cast alloys

Figure 4 summarizes the experimental CCT diagrams of Inconel 718 made by LPBF and suction casting after homogenization at 1180°C with different durations. NbC and δ are the two major precipitates observed in samples after cooling. Since the signal of the γ″/γ′ formation is undetectable in the dilatation curves, the CCT curves of γ″/γ′ are not given in Fig. 4. This is distinctly different from the work by Garcia et al. [14], in which multiple phases (i.e., the Laves phase, the M(C, N) phase, the δ phase, and the γ″/γ′ phases) were found to precipitate in sequence during cooling as shown in Fig. 1. One of the reasons for such a difference may be the large casting ingot (with a diameter of 530 mm) used in Garcia's work, of which the segregation of Nb should be much higher than that in the present work due to the slow cooling rate. The higher Nb segregation degree caused faster precipitation kinetics for the phases such as the γ″/γ′. The formation of the Laves phase during the cooling process in [14] remains a question because it usually forms during the solidification process [12,13] and no evidence of their formation during the solid-solid phase transformations has been reported to the best of the authors' knowledge.

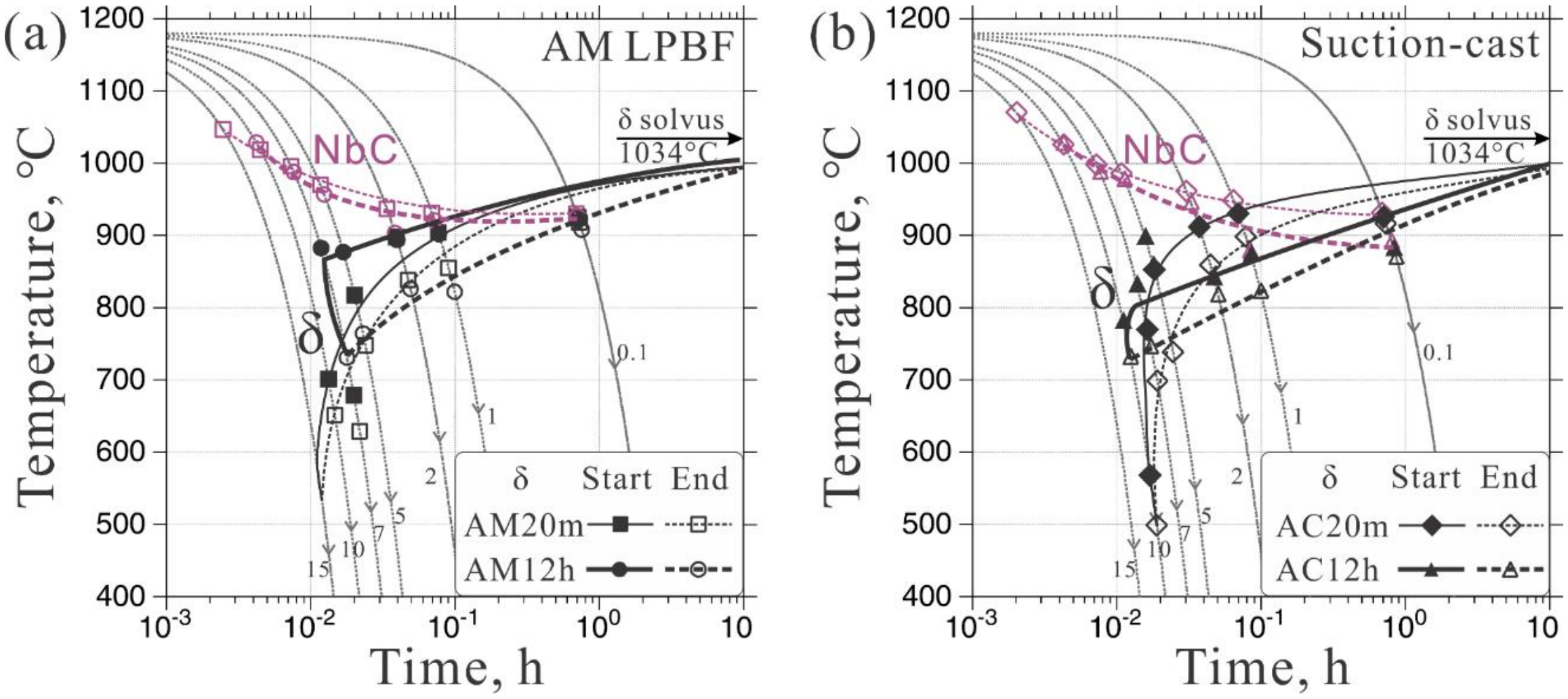

*Figure 4. Effects of homogenization on the CCT diagrams of (a) AM alloys homogenized for 20 min and 12 h; and (b) suction-cast alloys homogenized for 20 min and 12 h. The solvus temperature of the δ phase is predicted using the Thermo-Calc software TCNI8 database. Cooling rates in unit of K/s (from 0.1 to 15) are indicated in cooling curves superimposed to the CCT diagram plot.*

In this work, the solvus temperature of the δ phase in the suction-cast alloys is calculated to be 1034°C by the TCNI8 database, and it is used to represent the δ solvus temperature for both AM and suction-cast alloys as they have identical compositions. The solvus temperature of 1034°C is the formation temperature of the δ phase at equilibrium and thus is the highest limitation that the starting temperature of the δ phase can tend to approach at extremely slow cooling rate. Therefore, it is set as the limitation that the starting temperature of the δ phase can reach on the CCT diagrams (Figs. 4(a) & (b)).

According to Fig. 4, the phase transformation ending curves of NbC show a small difference in all CCT diagrams. These curves tend to lower temperature with a slower cooling rate because the longer heating durations from a slow cooling rate can promote the precipitation of NbC more sufficiently. However, a remarkable difference between CCT curves of δ phase in the alloys under different homogenization durations can be observed (Fig. 4). It should be pointed out that although the δ phase was observed in sample AM20m-15, no phase transformation signals can be detected by dilatometer, which can be because of the little phase fraction formed. Therefore, the CCT curve of alloy AM20m is extrapolated to the cooling rate of 15 K/s (Fig. 4(a)). Moreover, the manufacturing methods are also found to influence the CCT diagrams of the δ phase, as shown in Fig. 5. These findings indicate that the homogenization and manufacturing conditions can have significant effects on the phase transformation of the δ phase during continuous cooling. Such effects will be discussed in detail in the following sections.

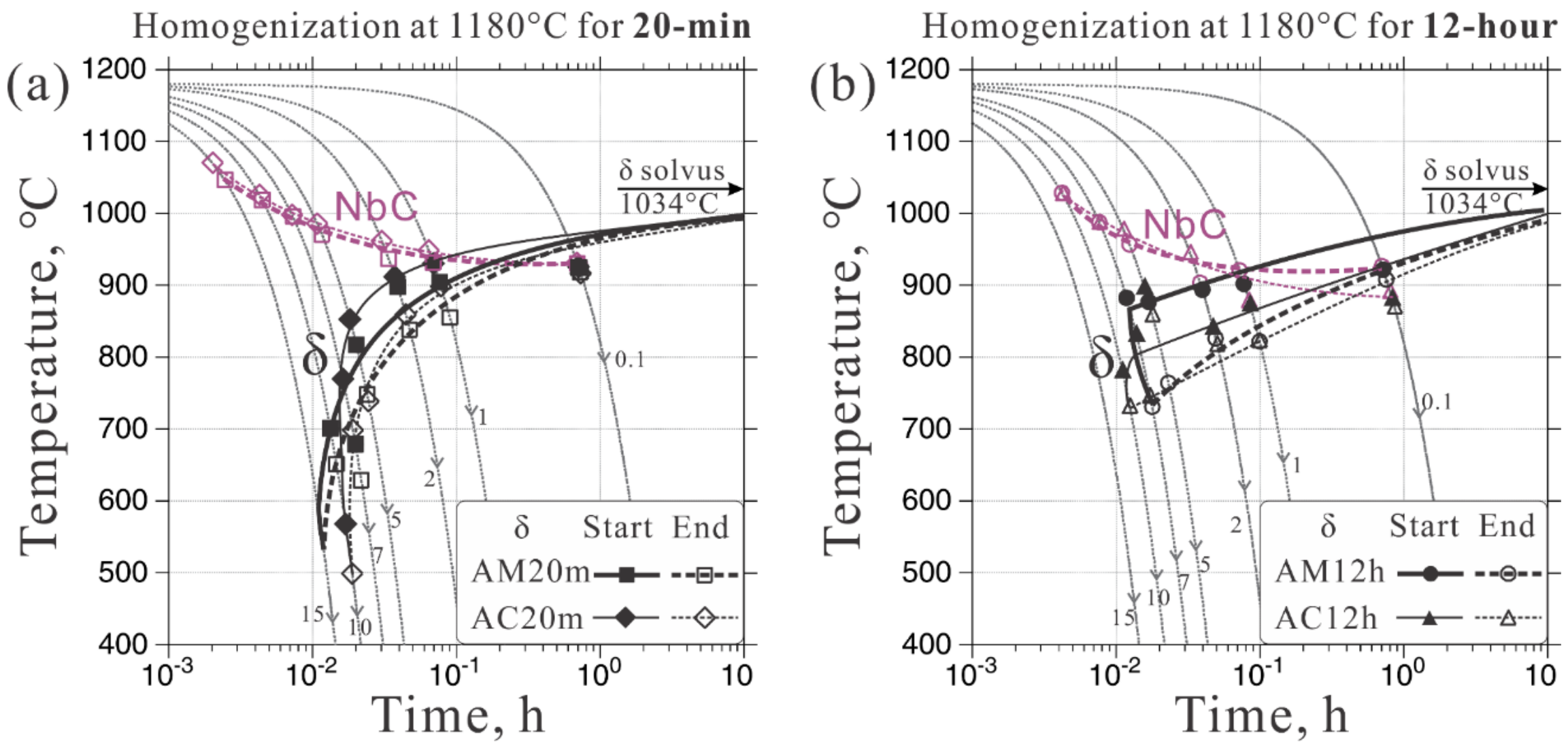

*Figure 5. Effects of manufacturing methods on the CCT diagrams of (a) AM & suction-cast alloys homogenized for 20 min; and (b) AM and suction-cast alloys homogenized for 12 h. The solvus temperature of the δ phase is predicted using the Thermo-Calc software TCNI8 database. Cooling rates in unit of K/s (from 0.1 to 15) are indicated in cooling curves superimposed to the CCT diagram plot.*

### 4.3. Effects of homogenization on the δ phase transformation upon cooling

#### *4.3.1. Nb homogenization in alloys*

According to Fig. 4, Table 3 summarizes the influence of homogenization on the phase transformation starting temperature, phase formation range, and critical cooling rate (as defined in Fig. 1(a)) of the CCT curves of the δ phase. Long (12-h) and short (20-min) homogenization times have opposite effects on the characteristics of CCT curves.

As summarized in Table 3 and Fig. 4(a), when comparing two AM alloys, i.e., AM20m and AM12h, the CCT curve of alloy AM20m has a lower starting temperature and a smaller formation range of the δ phase, but higher critical cooling rate. However, such effects of homogenization time on the starting temperature and the formation range of suction-cast alloys are reversed, as shown in Table 3 and Fig. 4(b). It should be noted that the difference in critical cooling rates between two suction-cast alloys, AC20m and AC12h, are found to be negligible based on the cooling rates tested in the current study.

Although the determined CCT diagrams by Garcia et al. [14] are distinctly different from this work, the homogenization condition clearly shows a significant impact on the solid-solid phase transformation during continuous cooling. In addition, Zhao et al. [13] found the Nb homogeneity degrees in the vicinity of NbC particles in Inconel 718 can vary with different manufacturing methods and homogenization conditions. The Nb homogeneity evolution was measured in [13] and replotted in Fig. 6. As shown in Fig. 6, for the AM alloys, the Nb homogeneity will decrease with extended homogenization durations (Figs. 6(a)&(b)), whereas in the suction-cast alloys, the Nb homogeneity will increase during homogenization with longer time (Figs. 6(c)&(d)). As the major element of the δ phase, Nb content and its homogeneity level in the matrix can affect the nucleation and precipitation kinetics of the δ phase.

**Table 3. Comparison of the effects of homogenization durations on the CCT characteristics of the δ phase.**

| Experimental Information | | Phase Formation and Growth Analysis | | | | CCT Diagram Characteristics | | |
|---|---|---|---|---|---|---|---|---|
| Sample* | Homogenization Time | Nb Homogeneity | Nucleation Potency | Growth Rate | Supercooling | Starting Temperature | Formation Range | Critical Cooling Rate ♦ |
| AM20m | 20-min | Higher + | Higher + | Lower – | Larger + | Lower – | Smaller – | Highest + |
| AM12h | 12-h | Lower – | Lower – | Higher + | Smaller – | Higher + | Larger + | Lowest – |
| AC20m | 20-min | Lower – | Lower – | Higher + | Smaller – | Higher + | Larger + | Medium ○ |
| AC12h | 12-h | Higher + | Higher + | Lower – | Larger + | Lower – | Smaller – | Medium ○ |

* AM20m and AM12h are homogenized samples manufactured by laser powder bed fusion, AC20m and AC12h are homogenized samples manufactured by suction casting.

♦ Critical cooling rate are rough values reading from the evaluated CCT diagram based on experimental data.

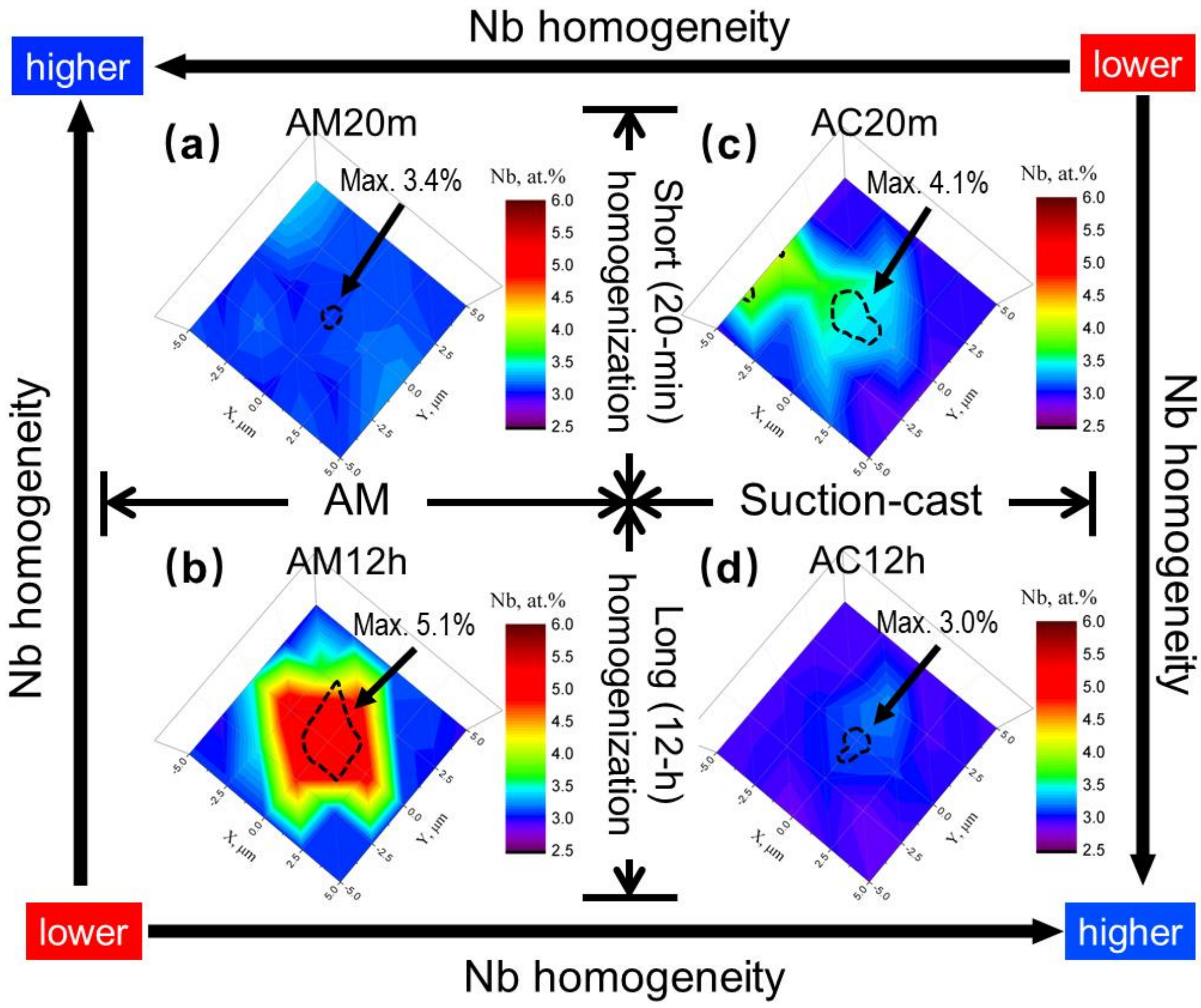

*Figure 6. Nb concentration contour diagrams indicating the Nb homogeneity in the vicinity of NbC carbides of different alloys after homogenization at 1180°C (AM: additive manufacturing, AC: suction casting). (a) AM20m, an alloy made by LPBF with homogenization for 20 min; (b) AM12h, an alloy made by LPBF with homogenization for 12 h; (c) AC20m, an alloy made by suction casting with homogenization for 20 min; (d) AC12h, an alloy made by suction casting with homogenization for 12 h. Dashed black squiggles profile the NbC carbides. The Nb concentration contour maps are developed using the EDS point identification with a grid size of 10 x10 μm.*

Figure 7 takes the comparison between alloys AM20m and AM12 as an example to illustrate the way Nb homogeneity affecting the precipitation kinetics. As shown in Fig. 7, the homogenized sample AM20m has higher Nb homogeneity than sample AM12h, thus each Nb atom in sample AM20m shares the same probability of becoming a potential nucleation site, which increases $N_0$, and the nucleation rate $N_r$ in Eq. (1) rises accordingly. Nevertheless, the relatively high Nb homogeneity around the nuclei leads $X_0^{Nb}$ in Eq. (2) to be low as the concentration fluctuation of Nb will be negligible. The supersaturation $\Delta X_0^{Nb}$ in Eq. (2) is hence reduced, resulting in a decrease in the growth rate. Therefore, the δ phase formation in alloy AM20m during continuous cooling primarily depends on the nucleation process while limited by the growth process. It is noteworthy that, some other defects, such as dislocations or stacking faults, can also act as the heterogeneous nucleation sites during isothermal phase transformations [38] for the precipitates since the nucleation barrier can be effectively reduced and the diffusion of the solute atoms can be promoted. However, in the present work, the continuous cooling processes offer the driving force for the nucleation and limit the diffusion of the atoms, which is different from the case in the isothermal phase transformations. Therefore, it can be presumed that the effect of defects on the phase transformations during continuous cooling processes is probably not as important as that in isothermal phase transformations, and it is reasonable to apply a homogenous nucleation mechanism for analysis. Nevertheless, the influence of defects on the athermal precipitation kinetics needs more dedicated study, while it is out of the scope of the current work.

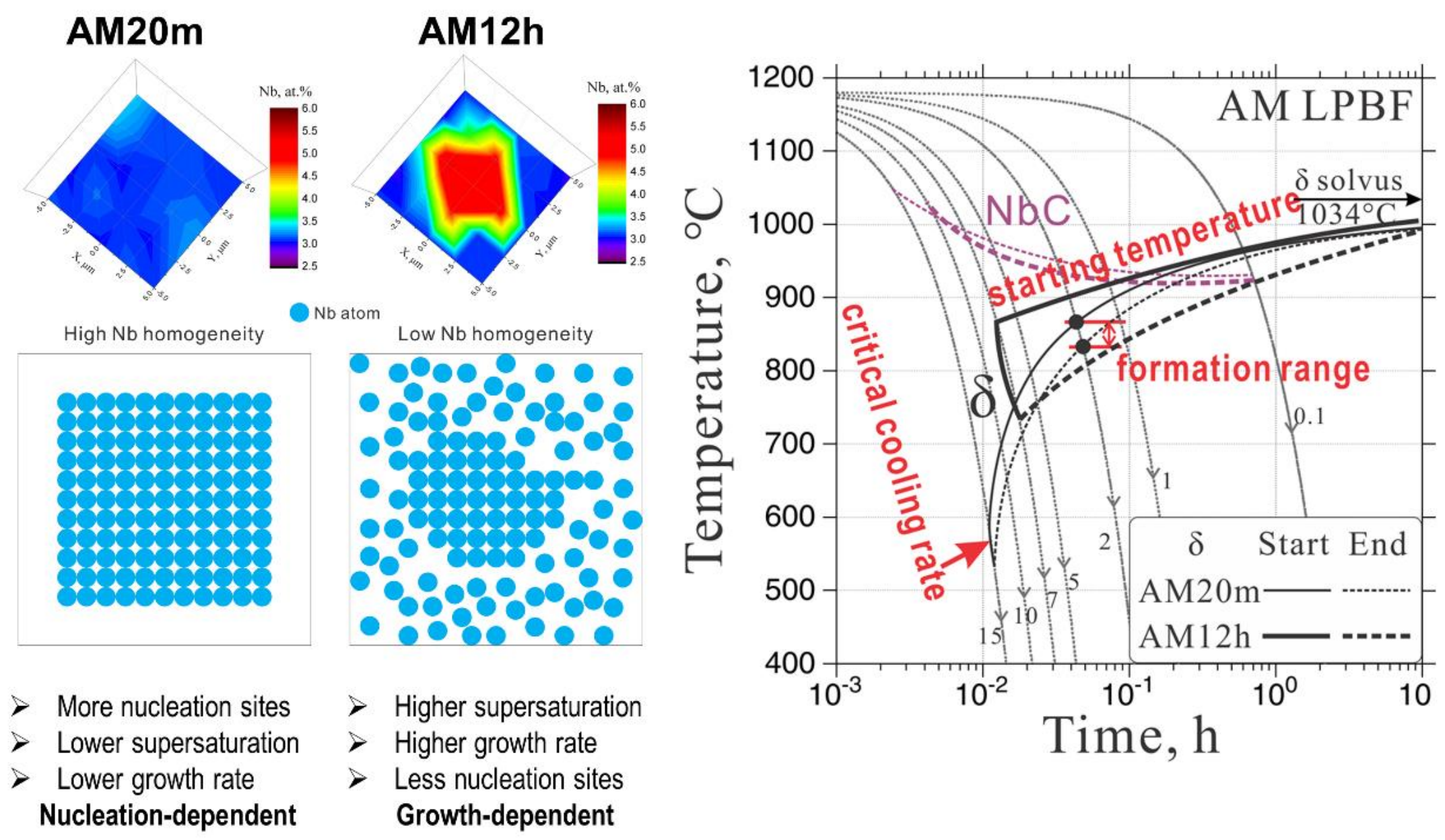

*Figure 7. Illustration of the effects of Nb homogeneity in the matrix near the NbC particles on precipitation kinetics of the δ phase (taking alloys AM20m and AM12h as an example).*

*4.3.2. Analysis based on classical nucleation and growth theory*

From Eq. (1), the nucleation rate $N_r$ can be significantly influenced by the nucleation barrier $\Delta G^*$, which can be reduced through obtaining an increased supercooling during the nonequilibrium cooling process. In alloy AM20m, the starting temperature of the CCT curve of δ is found to be lower than that in AM12h, as shown in Table 3 and the CCT diagram in Fig. 7. It indicates that the alloy AM20m has a larger supercooling degree during cooling when the δ phase starts to form, which is beneficial to the nucleation process. This may be because of the low Nb supersaturation in this alloy, with which a larger supercooling should be achieved to provide enough nucleation driving force to decrease the nucleation barrier. In addition, when the cooling rates for AM20m become faster, the supercooling of δ precipitation further increases with a higher driving force for nucleation. Because the precipitation in alloy AM20m mainly depends on the nucleation process, a higher cooling rate can hence promote nucleation by offering a higher nucleation driving force to reduce the nucleation barrier, making the nucleation easier. This allows the δ phase to form at high cooling rates, which explains the higher critical cooling rate observed for the δ phase in alloy AM20m (Table 3 and Fig. 7). In contrast, since the growth process of the δ phase is limited in alloy AM20m during cooling, the precipitation process finishes faster, leaving a relatively smaller formation range (Table 3 and Fig. 7).

In alloy AM12h, as can be seen in Fig. 7, due to the lower Nb homogeneity, the number of potential nucleation sites $N_0$ is less, as the Nb atoms close to the center of the Nb-rich area are more likely to become the nucleation sites. However, the local supersaturation around the nuclei can be higher. The growth process becomes the promoting factor of the precipitation, while the nucleation process limits the phase transformation. The precipitation kinetics of δ depends more on the precipitate growth in alloy AM12h during cooling. Under such circumstances, a small supercooling degree is beneficial to the diffusion of solute atoms during the growth process. Accordingly, in experiments, the CCT curve of the δ phase in alloy AM12h has a smaller supercooling, leading to a higher starting temperature comparing with that in alloy AM20m. The growth process in alloy AM12h can also last a longer time, and the formation range is thus larger (Table 3). In addition, when the cooling rates become higher in alloy AM12h, the phase transformation will be suppressed due to the diffusion of solute atoms is retarded, causing a lower critical cooling rate of 7 K/s (Table 3 and Fig. 7).

For the suction-cast alloys AC12h and AC20m, a similar explanation on the difference of the CCT curves can be applied. As can be seen in Figs. 6(c)&(d), Nb homogeneity increases with longer homogenization durations, so the alloy AC12h has a higher Nb homogeneity than alloy AC20m. According to the analysis above, for the suction-cast alloys homogenized for 20 min, the growth process dominates the precipitation process of δ phase, while for the suction-cast alloys homogenized for 12 h, the nucleation process dominates the precipitation kinetics of δ phase. This is opposite to the case in the AM alloys, as shown in Table 3.

*4.3.3. Analysis based on precipitation simulation*

Figure 8 is the simulated CCT diagram of the δ phase performed using TC-PRISMA. It shows that the low Nb homogeneity in the alloy, similar to AM12h, has higher starting temperatures and a lower critical cooling rate than the condition like AM20m with high Nb homogeneity. The simulation results agree reasonably well with the experimental observation shown in Fig. 4(a). This proves that the Nb homogeneity can affect the CCT diagrams of the δ phase in the way depicted by Fig. 7. It is noted that the difference of formation range observed in experiments (Fig. 4(a)) is not reflected by the simulation. This may be because the whole precipitation system in the precipitation simulation is considered as homogeneous (either for the case of 4.99 or 6.00 wt.% Nb). Therefore, the simulation cannot manifest the difference of diffusion processes caused by different Nb supersaturations between AM12h and AM20m. However, in practical, the lower Nb homogeneity in sample AM12h increases the supersaturation of Nb around NbC carbides, leading to a higher driving force and thus allowing Nb to diffuse more sufficiently at a relatively lower temperature to form δ.

In summary, with the same manufacturing method, the homogenization durations can affect the phase transformation behaviors of the δ phase during continuous cooling processes. This effect is achieved by varying the Nb homogeneity. For the AM alloys, extending the homogenization duration leads to the reduction of Nb homogeneity, making the precipitation kinetics of δ phase depends more on the growth process, yet for the suction-cast alloys, the extension of homogenization durations results in an increase in Nb homogeneity, and the precipitation kinetics mainly depends on nucleation process.

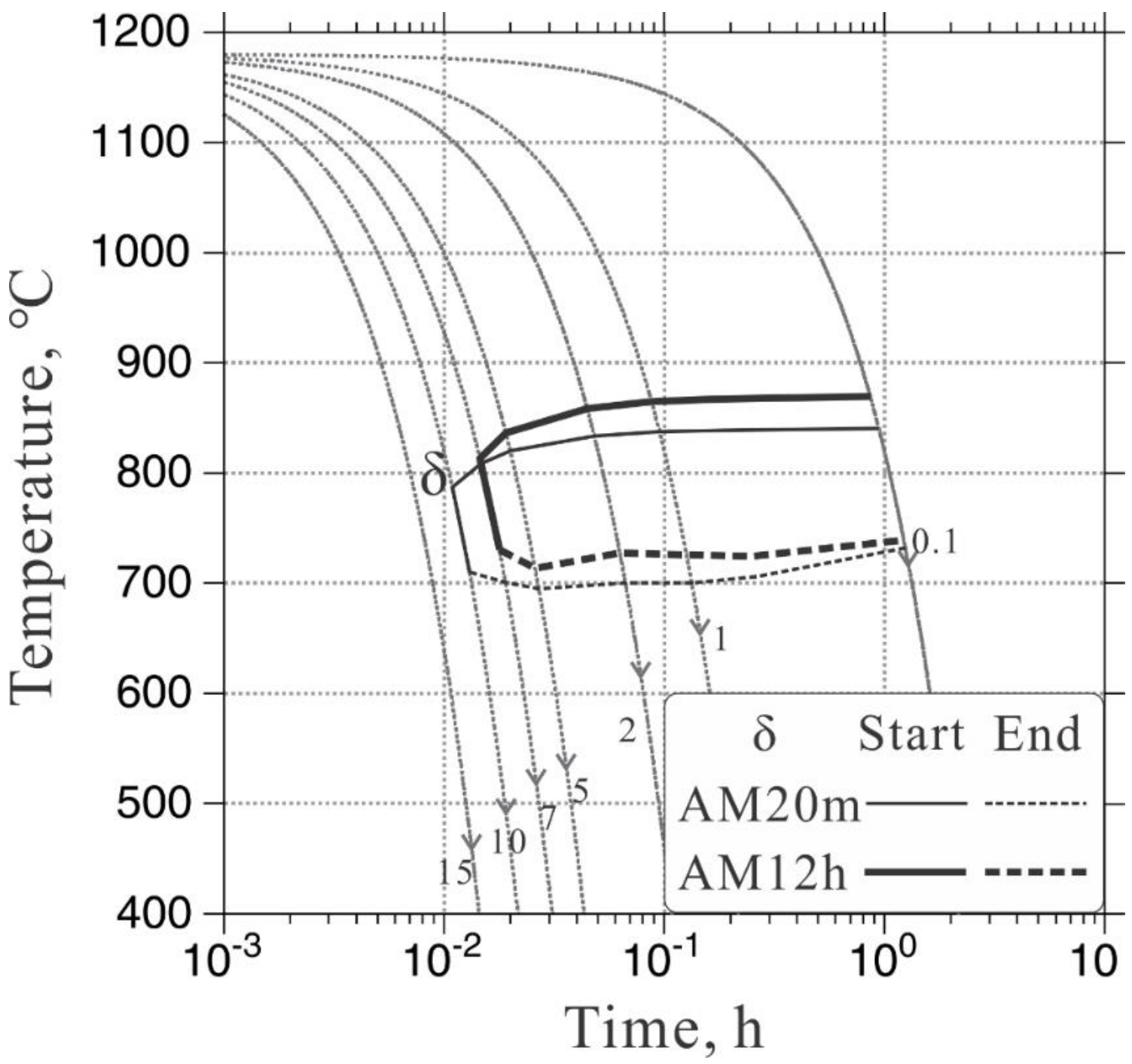

*Figure 8. Precipitation simulation of the δ phase for the qualitative validation of Nb homogeneity effects on the CCT diagrams. AM12h and AM20m represent the conditions of low Nb homogeneity and high Nb homogeneity, respectively.*

### 4.4. Effects of manufacturing methods on the CCT diagrams of the δ phase

The manufacturing methods are also found to influence the precipitation of the δ phase in the alloys subject to the same homogenization durations. As shown in Figs. 5(a)&(b) and listed in Table 3, with the same homogenization duration of 20 min (Fig. 5(a)), the CCT curve of δ phase in the AM alloy (AM20m) has a lower starting temperature, smaller formation range, but higher critical cooling rate compared with the suction-cast alloy (AC20m). However, under homogenization for 12 h (Fig. 5(b)), the AM alloy (AM12h) has a higher starting temperature, larger formation range, but a lower critical cooling rate than the suction-cast alloy (AC12h).

Such a difference can be readily explained through the Nb homogeneity analysis similar to Section 4.3. Figures 6(a)&(c) indicate alloy AM20m has a higher Nb homogeneity than alloy AC20m; therefore, it can be inferred that the nucleation process is the main factor contributing to the precipitation of δ phase in alloy AM20m. The supercooling degree in alloy AM20m should thus be large to provide sufficient nucleation driving force, which causes the reduction of starting temperature and increased critical cooling rate, as listed in Table 3. Meanwhile, the growth rate in AM20m is limited, attributed to a lower Nb supersaturation, which leads to the smaller formation range. The effects of manufacturing methods on the CCT curves of samples homogenized for 12 h listed in Table 3 can be interpreted in the same way.

In general, with the same homogenization durations, the manufacturing methods can influence the precipitation kinetics of the δ phase during continuous cooling by generating different Nb homogeneities. For the short duration of homogenization (20 min), the AM alloy has a higher Nb homogeneity, and the precipitation kinetics of the δ phase depends more on the nucleation process, yet the Nb homogeneity is lower in the suction-cast alloy, and the precipitation is more dependent on the growth process. The circumstances for long durations of homogenization (12 h) are reversed.

### 4.5. Effects of cooling rates on microhardness

Microhardness of alloys after continuous cooling is investigated by Vickers microhardness testing, and the results are presented in Fig. 9. For the tested alloys, as illustrated in Figs. 9(a)-(d), the cooling rates of 1-15 K/s result in comparable hardness values (229.5 $HV_{0.05}$ to 272.1 $HV_{0.05}$) with the as-homogenized samples. However, the hardness values achieved in the samples cooled at 0.1 K/s are evidently higher. Sample AM20m-01 exhibits the highest hardness of 421.1 $HV_{0.05}$, whereas the hardness values of the rest three samples cooled at 0.1 K/s vary from 349.8 $HV_{0.05}$ to 402.8 $HV_{0.05}$. This indicates some significant microstructure differences among these samples.

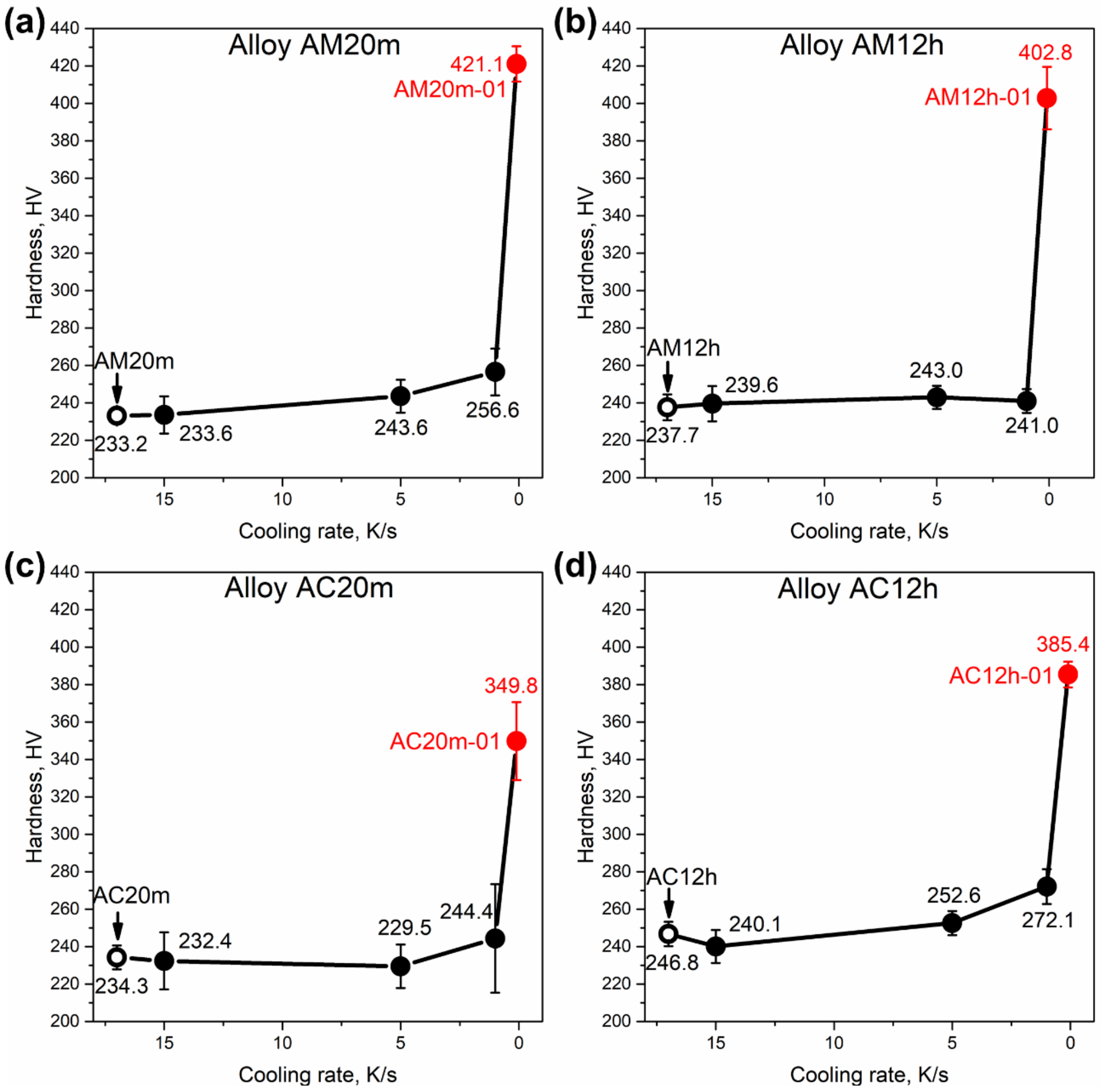

*Figure 9. Vickers hardness testing results of continuously cooled Inconel 718 samples after homogenization at 1180℃ with 20 min or 12 h: (a) alloy AM20m; (b) alloy AM12h; (c) alloy AC20m; (d) alloy AC12h. AC means the alloys are prepared using suction casting and AM is noted for the samples prepared by LPBF.*

Since the samples cooled at 0.1 K/s all have relatively high hardness (Fig. 9) compared to other samples cooled at higher cooling rates, these samples are etched for microstructure observation. In the etched microstructures, a large number of δ precipitates can be found to form along grain boundaries in all these samples, as seen in the subfigures of Figs. 10(a)-(d). Moreover, the appearance of subgrain boundaries is observed in all samples with a cooling rate of 0.1 K/s (Fig. 10). Such subgrain boundaries are not observed in the samples cooled at higher rates, e.g., in samples AC20m-5 and AM20m-5 (Fig. SI-1), which implies they may be related to the high hardness in the samples cooled at 0.1 K/s. The subgrains probably form through the tangling of dislocations, which are generated through the loss of coherency of δ/γ phase boundaries during the precipitation of the δ phase [39,40]. The δ particles can further pin the subgrain boundaries and regard their movement. As a result, small subgrains form due to the δ precipitation, and the hardness is elevated accordingly. Conversely, precipitation of the δ phase is more limited at cooling rates faster than 0.1 K/s, so the subgrains are less likely to form. This explains the lower hardness obtained in the rest as-cooled samples.

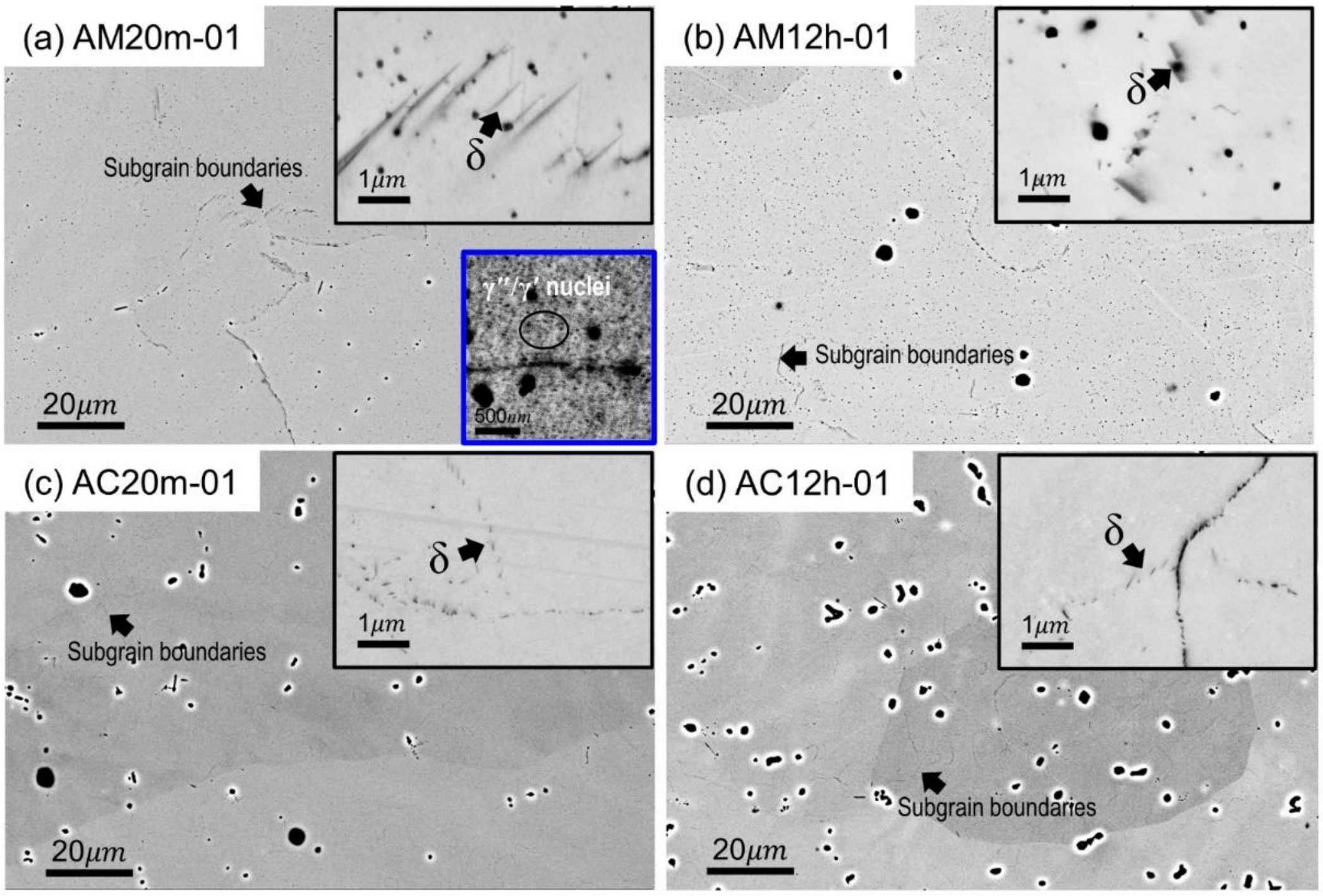

*Figure 10. SEM micrographs on etched Inconel 718 samples cooled at 0.1 K/s: (a) sample AM20m-01; (b) sample AM12h-01; (c) sample AC20m-01; (d) sample AC12h-01.*

Moreover, the precipitation strengthening due to γ″/γ′ formation could be one of the reasons that cause the sample AM20m-01 to have the highest hardness among all the samples cooled at 0.1 K/s (Fig. 9). As shown in Fig. 10(a), a trace of γ″/γ′ nuclei is observed in sample AM20m-01, although their dilatation signals may be too weak to be detected. Geng et al. [23] reported that the γ″ precipitation was observed at cooling rates between 0.1-20 K/min (0.0017-0.33 K/s), which is consistent with the observation in sample AM20m-01. However, in Figs. 10(b)-(d), no γ″/γ′ nuclei can be observed in samples AM12h-01, AC20m-01, and AC12h-01. Hence, the occurrence of γ″/γ′ nuclei can explain the highest hardness in sample AM20m-01 among all samples cooled at 0.1 K/s. The exclusive precipitation of γ″/γ′ in sample AM20m-01 is probably due to the homogenized alloy AM20m has the largest amount of Nb dissolved into the matrix, since the least amount of Nb-rich phases, i.e., Laves_C14 and NbC carbides, was found in this alloy [13]. Consequently, it leads to a higher Nb concentration in the γ matrix of sample AM20m-01 compared with other homogenized alloys shown in Fig. 10 and makes the γ″/γ′ precipitate readily.

## 5. Conclusions

(1) The CCT diagrams of AM and suction-cast Inconel 718 alloys after homogenizations at 1180°C for 20 min and 12 h are established based on microstructure characterization and dilatometry analysis. Both NbC carbide and δ phase are found to precipitate during the continuous cooling processes in all alloys at appropriate cooling rates.

(2) Homogenization durations as well as manufacturing methods can affect the phase transformation behaviors, i.e., the starting temperature, the precipitation formation range, and the critical cooling rate of δ phase during cooling, by changing the Nb homogeneity in the matrix near NbC particles. The precipitation kinetics of the δ phase during cooling depends more on the nucleation process in alloys with higher Nb homogeneity, while the growth process is predominant in alloys with lower Nb homogeneity.

(3) Compared with the samples cooled at higher rates, 0.1 K/s cooling rate can achieve the highest hardness in all alloys with different homogenization durations and manufacturing conditions. This is probably due to the formation of subgrains as a result of the abundant precipitation of the δ phase. Compared with other samples cooled at 0.1 K/s, the hardness of the sample AM20m-01 with higher Nb homogenization is further increased due to the presence of γ″/γ′ nuclei.

(4) This work indicates that the ad-hoc design of post-heat treatment for microstructure engineering of AM alloys becomes essential. A careful analysis of microstructure influenced by cooling processes is critical for the heat treatment design of AM alloys.

(5) Generally, this study implies that homogenization and manufacturing conditions can influence phase transformations during continuous cooling in alloys with more prominent elemental microsegregation and sluggish diffusion. Such effects can be achieved by the modification of elemental homogeneity in the alloys, which will further affect the kinetics of nucleation and growth processes. However, such effects are presumed not to be critical in alloys with fast diffusion or diffusionless phase transformations, more insightful investigations regarding the impact factors on continuous cooling process of these alloys are desired.

(6) During phase transformation modeling, the CCT diagram may not be the best choice to integrate with other thermal modeling simulations to understand the microstructure evolution of Inconel 718 alloys in additive manufacturing with cyclic heating and cooling.

## CRediT author statement

Yunhao Zhao: Validation, Formal analysis, Investigation, Data Curation, Writing - Original Draft; Liangyan Hao: Validation, Investigation, Writing - Review & Editing; Qiaofu Zhang: Validation, Investigation, Writing - Review & Editing; Wei Xiong: Conceptualization, Methodology, Investigation, Writing - Review & Editing, Supervision, Project administration, Funding acquisition.

## Data availability

The raw/processed data required to reproduce these findings cannot be shared at this time as the data also forms part of an ongoing study.

## Declaration of Competing Interest

The authors declare that they have no known competing financial interests or personal relationships that could have appeared to influence the work reported in this paper.

## Acknowledgment

The authors thank the National Aeronautics and Space Administration for the financial support under the Grant Number (NNX17AD11G). Authors are also grateful for the Thermo-Calc company on the software and databases provided for CALPHAD modeling through the ASM Materials Genome Toolkit Award. Ms. Yinxuan Li is appreciated for the help of sample preparation under the support through the Mascaro Center for Sustainable Innovation at the University of Pittsburgh.